\documentclass[twocolumn,prl]{revtex4}
\usepackage{amssymb,bbm,bm}
\usepackage{graphicx}
\usepackage{bm}
\usepackage{amsmath}
\usepackage[usenames]{color}
\usepackage{epsfig}
\usepackage{hyperref}
\usepackage{subfigure}
\usepackage[sans]{dsfont}
\hypersetup{colorlinks=true, citecolor=blue,
linkcolor=blue,urlcolor=blue }

\begin{document}

\title{Localization in a one-dimensional alloy with an arbitrary distribution of spacing between impurities: Application to L\'{e}vy glass}
\author{Reza Sepehrinia}\email{sepehrinia@ut.ac.ir}
\affiliation{Department of Physics, University of Tehran, Tehran 14395-547, Iran}

\begin{abstract}
We have studied the localization of waves in a one-dimensional lattice consisting of impurities where the spacing between consecutive impurities can take certain values with given probabilities. In general, such a distribution of impurities induces correlations in the disorder. In particular with a power-law distribution of spacing, this system is used as a model for light propagation in L\'{e}vy glasses. We introduce a method of calculating the Lyapunov exponent which overcomes limitations in the previous studies and can be easily extended to higher orders of perturbation theory. We obtain the Lyapunov exponent up to fourth order of perturbation and discuss the range of validity of perturbation theory,  transparent states, and anomalous energies which are characterized by divergences in different orders of the expansion. We also carry out numerical simulations which are in agreement with our analytical results.
\end{abstract}
\maketitle

\section{Introduction}

Propagation of waves in random media is a subject of interest in a variety of areas ranging from visible light propagation in human tissue for medical purposes to electromagnetic wave propagation in interstellar clouds \cite{van1999multiple,ishimaru1978wave,maynard2001acoustical,belitz1994anderson}. A daily life example is the propagation of sunlight in clouds. Even though the sun can be invisible on a cloudy day we still can see diffuse light coming in all directions. Usually, the propagation of light through such a scattering medium can be approximated by normal diffusion. Likewise the transport of heat or sound waves, in certain length scales, is also described by normal diffusion.

In a recent study, a disordered optical medium has been designed in which the propagation of light is governed by superdiffusion rather than normal diffusion \cite{barthelemy2008levy}. These engineered materials, which are named L\'{e}vy glasses, are realized with an assembly of transparent microspheres, with a controlled size distribution, embedded in a scattering medium. The power-law distribution of the size of these microspheres induces a heavy tail distribution of the step length for light rays and therefore a L\'{e}vy-type random walk through the medium.  Although the transport properties can be described by the random walk model to a large extent, it would be interesting to see how these properties are influenced by wave phenomena such as interference.

This problem is addressed in Ref. \cite{zakeri2015localization} where a one-dimensional discrete model of a mechanical analog of L\'{e}vy glass, represented by a harmonic chain of coupled oscillators, is studied. Randomness is introduced in the spacing between impurities with a power-law distribution $p(s)\propto s^{-(\alpha+1)}$. It turns out that this model exhibits anomalous localization properties. Namely, the localization length of vibrational modes at low frequencies (long wavelength) exhibits a scaling behavior $\xi\propto \omega^{-\alpha}$, in contrast with the standard scaling behavior $\xi\propto \omega^{-2}$ for uncorrelated disorder. Similar scaling behavior is found in a continuous model of a one-dimensional layered system at long-wavelength limit \cite{asatryan2018anderson}. In Ref. \cite{herrera2019heat} the discrete model is reconsidered and an analytical formula for the power spectrum of the mass distribution of this model is obtained.

The studies on the discrete model \cite{zakeri2015localization,herrera2019heat} rely on the second-order perturbative expression for the Lyapunov exponent which is obtained \cite{izrailev1999localization} in terms of the correlation functions of the disorder. For the L\'{e}vy-type distribution of impurities, this method does not lead to a conclusive result in the entire range of the power-law exponent $\alpha$. Here, we introduce an alternative method of calculating the Lyapunov exponent for this model which does not require determining the correlation functions. Our approach provides a systematic way of calculating the higher orders of perturbation expansion with any given distribution of spacing between impurities. In this paper, we study the problem of electron localization \cite{falceto2011conductance,wells2008quantum,iomin2009lyapunov} although mathematically it is equivalent to the problem of mechanical vibrations.

Higher-order terms in the expansion allow us to study the phenomenon of the Kappus-Wegner anomaly \cite{kappus1981anomaly} which is the result of constructive interference of certain scattering amplitudes \cite{alloatti2008anomalies} and characterized by the enhancement of the localization length at certain isolated energies. It turns out that for the special random potential that we study here, such anomalies occur at several energies, which is in contrast with the white noise potential. We also investigate the range of validity of the perturbative expansion, transparent states, and carry out numerical simulations and compare them with our analytical results.
\\

\section{Model}

\begin{figure}[t]
\epsfxsize9truecm \epsffile{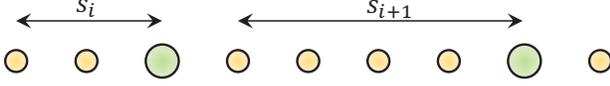} \caption{(Color online) Schematic illustration of the tight-binding chain. Small circles show the background lattice with zero on-site potential and the large circles are the impurities with on-site potential $\lambda U$. Here, $s_i=3$ and $s_{i+1}=5.$}\label{chain}
\end{figure}

The model under consideration is a one-dimensional tight-binding chain (Fig. \ref{chain}), represented by the discrete Schr\"{o}dinger equation
\begin{equation}\label{model}
   \Psi_{n+1}+\Psi_{n-1}+\lambda U_n\Psi_n = E\Psi_n.
\end{equation}
The potential $U_n$ is assumed to take two values $U_n=0$ and $U_n=U$ in the following way. There are sequences of $U_n=0$ with length $s-1$ and after each such sequence there will be an impurity with $U_n=U$. The sequence length $s$ is a random variable with integer values $s=1,2,\dots$, drawn from a given distribution $p(s)$.

As we mentioned, vibrations of atoms in a one-dimensional crystal with harmonic forces between nearest-neighbor atoms and binary mass distribution can also be described with this model. The following replacement should be done, $E-\lambda U_n=2-m_n\omega^2/\kappa$, with $\omega$ being the frequency and $\kappa$ being the spring constant. For the binary mass distribution $m,M$, we can use $E=2-m\omega^2/\kappa$ and $\lambda U=(M-m)\omega^2/\kappa$ to transform the mass-spring model to model Eq. (\ref{model}).

\section{Perturbation theory}

The solution of Eq. (\ref{model}), in the presence of the weak random potential, can be treated perturbatively \cite{thouless1974electrons,balian1983ill,derrida1984lyapounov,sepehrinia2010irrational,sepehrinia2021localization} by rewriting it in terms of variables $R_n=\frac{\Psi_{n+1}}{\Psi_n}$,
\begin{equation}\label{EqR}
     R_n=E-\lambda U_n-\frac{1}{R_{n-1}}.
\end{equation}
The solution of Eq. (\ref{EqR}) for nonzero weak random potential can be expressed as the following expansion in powers of disorder strength,
\begin{equation}\label{Rexpansion}
     R_n=A \exp(B_n \lambda + C_n \lambda^2 + D_n \lambda^3 + F_n \lambda^4 +\cdots).
\end{equation}
By substituting (\ref{Rexpansion}) in (\ref{EqR}) and collecting terms in different orders of $\lambda$, one gets \cite{gardner1984laplacian}
\begin{subequations}\label{ABC}
\begin{align}\noindent
&A^2+1= AE, \label{eq-a}\\
&A^2B_n = B_{n-1} - A U_n, \label{eq-B} \\
&A^2(C_n + \tfrac{1}{2} B_{n}^2) = C_{n-1} - \tfrac{1}{2} B_{n-1}^2. \label{eq-C}\\
&A^2(D_n + B_nC_n + \tfrac{1}{6} B_n^3) = D_{n-1} - B_{n-1}C_{n-1} + \tfrac{1}{6} B_{n-1}^3 \nonumber \\ \label{eq-D}\\
&A^2(F_n + B_nD_n + \tfrac{1}{2} C_n^2 + \tfrac{1}{2} B_n^2C_n + \tfrac{1}{24}B_n^4) = \nonumber \\ \label{eq-E}
& \ \ \ F_{n-1} - B_{n-1}D_{n-1} - \tfrac{1}{2} C_{n-1}^2 + \tfrac{1}{2} B_{n-1}^2C_{n-1} - \tfrac{1}{24}B_{n-1}^4 \nonumber \\
\end{align}
\end{subequations}
The rate of exponential growth of solutions, i.e., the inverse localization length, is determined by the Lyapunov exponent, which is given by
\begin{eqnarray}\label{LE}
    \gamma(E)&=&\lim_{N\rightarrow \infty}\frac{1}{N}\sum_{n=1}^N \log R_n=\langle \log R\rangle\\
    &=&\log A +\lambda \langle B\rangle+\lambda^2\langle C\rangle+\lambda^3\langle D\rangle+\cdots.
\end{eqnarray}
The averages in the last equation can be obtained using the average of Eqs. (\ref{eq-B}-\ref{eq-E}) and their multiplications. We have
\begin{eqnarray}\label{avB1}
  \langle B\rangle &=& -\tfrac{1}{A-A^{-1}}\langle U \rangle, \\
  \langle C\rangle &=& - \tfrac{1}{2}\tfrac{A+A^{-1}}{A-A^{-1}}\langle B^2\rangle, \label{avC1} \\
  \langle D\rangle &=& - \tfrac{A+A^{-1}}{A-A^{-1}} \langle BC\rangle - \tfrac{1}{6}\langle B^3\rangle\label{avD1} \\
  \langle F\rangle &=& - \tfrac{A+A^{-1}}{A-A^{-1}} (\langle BD\rangle+\tfrac{1}{2}\langle C^2\rangle+\tfrac{1}{24}\langle B^4\rangle) - \tfrac{1}{2}\langle B^2C\rangle. \label{avE1}\nonumber \\
\end{eqnarray}
There is an obstacle in calculating the averages on the right-hand side. Unlike the uncorrelated case \cite{derrida1984lyapounov}, the averages such as $\langle X_n^p U_n^q\rangle$ can not be replaced with $\langle X_n^p\rangle \langle U_n^q\rangle$ because of the correlations in the potential. Here $X_n$ can be $B_n,C_n,B_nC_n$ etc. In order to calculate such averages, since the process $X_n$ is stationary, we can consider them as spatial averages
\begin{equation}\label{}
     \langle X_n^p U_n^q\rangle = \tfrac{1}{L}\sum_n X_n^p U_n^q,
\end{equation}
and also using the fact that the potential $U_n$ is only nonzero on impurities we will have
\begin{equation}\label{}
     \langle X_n^p U_n^q\rangle = \tfrac{1}{L} U^q \sum_i X_i^p,
\end{equation}
where the sum is restricted to impurity positions. Since $L=N \langle s\rangle$ with the $N$ being the average number of impurities in the sequence of length $L$ we will have
\begin{equation}\label{}
     \langle X_n^p U_n^q\rangle = \tfrac{U^q}{\langle s\rangle}  \langle X^p\rangle_{\text{imp}},
\end{equation}
where $ \langle \cdot \rangle_{\text{imp}}$ means the average of the values on the impurities.
As an example, the average of the potential is given by
\begin{eqnarray}\label{avU}
 \langle U^q \rangle &=& \tfrac{U^q}{\langle s \rangle},
\end{eqnarray}
In order to calculate the averages such as $\langle X^p\rangle_{\text{imp}}$, we need to know the values of $X$ on impurity positions which we indicate by index $i$. First, we derive the recurrence relations in terms of the impurity index. As it is considered the number of lattice points between two consecutive impurities is $s_i-1$, by starting with $B_{i-1}$ and applying $B_n=A^{-2}B_{n-1}$, $s_i-1$ times followed by $B_n=A^{-2}B_{n-1}-A^{-1}U$ for the last point (which is an impurity) we obtain $B_i$
\begin{eqnarray}\label{eq-Bd}
  B_{i}=A^{-2s_i} B_{i-1} - A^{-1}U.
\end{eqnarray}
The advantage of the above recursive relation is that we can now use the statistical independence of $B_{i-1}$ and $s_i$ because $s_i$'s are assumed to be uncorrelated. By taking the average of both sides and using the statistical independence i.e. $\langle A^{-2s_i} B_{i-1}\rangle_{\text{imp}}=\langle A^{-2s_i}\rangle \langle B_{i-1}\rangle_{\text{imp}}$ we obtain
\begin{eqnarray}
  \langle B \rangle_{\text{imp}}=\tfrac{A^{-1}U}{\langle A^{-2s} \rangle -1},
\end{eqnarray}
where $\langle A^{-2s} \rangle=\sum_s p(s)A^{-2s}$. We use the calligraphic font to indicate the averages  $\mathcal{A}_m=\langle A^{-ms} \rangle$, for convenience. We emphasize the difference between this average and the total average of $B_n$, given by Eq. (\ref{avB1}) which by inserting from Eq. (\ref{avU}) is as follows
\begin{eqnarray}\label{avB}
  \langle B \rangle=-\tfrac{U}{\langle s\rangle(A -A^{-1})}.
\end{eqnarray}
Similarly, we can calculate $\langle B^2 \rangle_{\text{imp}}$ and then $\langle B^2 \rangle$
\begin{eqnarray}
&&  \langle B^2 \rangle_{\text{imp}} = \tfrac{(1+\mathcal{A}_2)U^2}{A^2(1-\mathcal{A}_2)(1-\mathcal{A}_4)}, \\
&&  \langle B^2 \rangle = \tfrac{(1+\mathcal{A}_2)U^2}{\langle s \rangle(A^2-A^{-2})(1-\mathcal{A}_2)},
\end{eqnarray}
from which we obtain
\begin{equation}\label{avC}
\langle C \rangle = -\tfrac{(1+\mathcal{A}_2)U^2}{2\langle s \rangle(A-A^{-1})^2(1-\mathcal{A}_2)}.
\end{equation}
By taking the real part of the above expression we reproduce the second-order result of Ref. \cite{herrera2019heat}
\begin{equation}\label{reC}
\text{Re} \ \langle C \rangle = -\tfrac{U^2}{2\langle s \rangle(A-A^{-1})^2}\tfrac{1-|\mathcal{A}_2|^2}{|1-\mathcal{A}_2|^2}.
\end{equation}
where we have assumed the limit $A\rightarrow e^{ik}$ to be taken. The wave vector $k$ is related to the energy via the zero-order equation (\ref{eq-B}), which gives the dispersion relation $E=2\cos k$.

Having introduced the approach we can now proceed to calculate the higher orders of perturbation.
The final result up to fourth-order is obtained as follows
\begin{eqnarray}\label{gamma}
 \gamma &=& \log A + \lambda \tfrac{U}{\langle s\rangle(A - A^{-1})} \nonumber \\
 &&- \frac{\lambda^2}{2} \tfrac{(1+\mathcal{A}_2)U^2}{\langle s  \rangle(A-A^{-1})^2(1-\mathcal{A}_2)}
 - \frac{\lambda^3}{3} \tfrac{(1+4\mathcal{A}_2+\mathcal{A}_2^2)U^3}{\langle s \rangle(A-A^{-1})^3(1-\mathcal{A}_2)^2} \nonumber \\ &&-\frac{\lambda^4}{4} \tfrac{(1+\mathcal{A}_2)(1 + 8 \mathcal{A}_2 - \mathcal{A}_2^2 + \mathcal{A}_4 - 8 \mathcal{A}_2 \mathcal{A}_4 -\mathcal{A}_2^2 \mathcal{A}_4 )U^4}{\langle s \rangle(A-A^{-1})^4(1-\mathcal{A}_2)^3(1-\mathcal{A}_4)}\nonumber \\
 &&+O(\lambda^5).
\end{eqnarray}
\section{Transparent states}

The simplest choice for the distribution of sequence length $s$ is $p(s)=\delta_{s,s_0}$. This means we will have one repeating pattern therefore a periodic potential. Thus all states are expected to be Bloch wave functions that are delocalized i.e. the real part of the Lyapunov exponent must be zero. We can see that this is the case in our result by noting that $\mathcal{A}_2= A^{-2s_0} $  which has a unit modulus, and as a result, the real part of each term in the expansion Eq. (\ref{gamma}) vanishes identically. This should be the case in all orders of perturbation because, as we mentioned, the potential is periodic in this case. Although the potential is periodic, since there are two types of atoms in the chain, the energy band splits into $s_0$ subbands with $s_0-1$ gaps. This latter property will be relevant in our later discussion.

If there is more than one value of $s$ then the chain will be disordered. Even in such a random case, there could be some energies where the corresponding states are fully transparent. Similar states is known for example in aperiodic Kronig-Penney \cite{izrailev2001mobility} and the random dimer \cite{izrailev1999localization} models. The present model may also possess such states which we demonstrate with a couple of examples. Let us take $p(s)=p \delta_{s,s_1}+ q \delta_{s,s_2}$ where $p+q=1$. We need $\mathcal{A}_2= p e^{-2iks_1}+q e^{-2iks_2}$ to have unit a modulus but $\mathcal{A}_2\neq 1$. The only solution for $|p e^{-2iks_1}+q e^{-2iks_2}|=1$ such that $p e^{-2iks_1}+q e^{-2iks_2}\neq 1$ is $k=n\pi/(s_2-s_1)$. This can be interpreted using the fact that the difference in the phase acquired by the solutions of Eq. (\ref{model}) between two consecutive impurities will be zero or $\pi$ if $e^{ik(s_2-s_1)}=\pm1$. Therefore the incoming wave perceives a periodic potential and will be a Bloch wave with an overall phase of zero or $\pi$.

This can also be seen if Eq. (\ref{model}) is expressed in terms of the transfer matrices
\begin{equation}\label{TM}
\left(\begin{array}{c}
\Psi_{n+1}  \\
\Psi_{n}
\end{array}\right)=
\left(\begin{array}{cc}
E-\lambda U_n & -1 \\
1 & 0
\end{array}\right)
\left(\begin{array}{c}
\Psi_{n}  \\
\Psi_{n-1}
\end{array}\right).
\end{equation}
The transfer matrix of the chain will be as
\begin{equation}\label{}
\cdots
\left(\begin{array}{cc}
E & -1 \\
1 & 0
\end{array}\right)^{s_1}
\left(\begin{array}{cc}
E-\lambda U & -1 \\
1 & 0
\end{array}\right)
\left(\begin{array}{cc}
E & -1 \\
1 & 0
\end{array}\right)^{s_2}
\cdots.
\end{equation}
If we have
\begin{equation}\label{tm1}
\left(\begin{array}{cc}
E & -1 \\
1 & 0
\end{array}\right)^{s_2-s_1}=\pm\textbf{1},
\end{equation}
then the transfer matrix of the whole chain will be the same as that of a periodic chain up to a sign. This means that the corresponding state will be a Bloch wave. The solutions of the Eq. (\ref{tm1}) are the same as we obtained above.

Similarly, for the probability distribution $p(s)=p \delta_{s,s_1}+ q \delta_{s,s_2} + r \delta_{s,s_3}$ with $p+q+r=1$, the wave vectors that correspond to transparent states, are common multiples of
$\pi/(s_2-s_1)$ and $\pi/(s_3-s_1)$ i.e. $k=n\pi/(s_2-s_1)=m\pi/(s_3-s_1)$ where $n$ and $m$ are integers.

\section{Validity of perturbative expansion}
\subsection{Band edges}
As in the case of uncorrelated disorder \cite{derrida1984lyapounov} we see that if $A\rightarrow \pm1$  i.e. $E\rightarrow \pm2$, each term of the expansion Eq. (\ref{gamma}) diverges because of the factor $(A-1/A)^n$ in the denominator. This is due to the fact that the expansion in integer powers of $\lambda$ is incorrect. It turns out that for an uncorrelated disorder with zero mean in the neighborhood of the band edges $\gamma\propto \lambda^{2/3}$ \cite{derrida1984lyapounov}. However in addition to the term $(A-1/A)^n$ in the denominator the factor $(1-\mathcal{A}_2)^{n-1}$, which comes from correlated disorder, also becomes zero in this limit. This can change the scaling of the Lyapunov exponent with disorder strength at the band edge. The particular potential that we have used has a nonzero mean value therefore the Lyapunov exponent scales as $\gamma\propto \lambda^{1/2}$ at the band edge.

In addition to band edges, depending on the distribution $p(s)$, the factor $(1-\mathcal{A}_2)^{n-1}$ causes other divergences which are associated with the above mentioned energy gaps inside the band. Again let us look at the binary distribution $p(s)=p \delta_{s,s_1}+ q \delta_{s,s_2}$ for which $\mathcal{A}_2=p e^{-2iks_1}+q e^{-2iks_2}$. If the $s_1, s_2$ and $k$ are such that $e^{-2iks_1}=e^{-2iks_2}=1$ then the expansion will diverge at the corresponding energy. Below we will illustrate this in the figures.

\subsection{Anomalous energies}
The fourth-order term in the expansion has $1-\mathcal{A}_4$ in the denominator which also makes this term diverge at certain points. This is similar to what happens at the band center of the uncorrelated disorder model which is known as the Kappus-Wegner anomaly that is characterized with an enhancement in the localization length due to the constructive interference of certain scattering amplitudes. In our case, there could be multiple anomalous energies of this type at which the fourth-order term diverges.

At the band center ($k=\pi/2$) we have $\mathcal{A}_4=\sum_s p(s)e^{-2i\pi s} =1$. Therefor close to the band center the fourth order term satisfies
\begin{equation}\label{4th}
 (1-\mathcal{A}_4)\langle F\rangle \simeq  \tfrac{(1+\mathcal{A}_2)^2 U^4}{32\langle s \rangle(1-\mathcal{A}_2)^2}.
\end{equation}
If $\mathcal{A}_2\neq 1$ the band center anomaly exists unless $\mathcal{A}_2\rightarrow -1$ in which case the fourth-order term (as well as other terms) would be zero and we will have a transparent state at $E=0$ as we mentioned earlier.

Unlike the uncorrelated disorder case, here the fourth-order term may diverge at other energies too, because $1-\mathcal{A}_4$ may have other roots which give rise to similar anomalies out of the band center. For example in the binary distribution the roots of $\mathcal{A}_4=p e^{-4iks_1}+q e^{-4iks_2}=1$ are $k=n\pi/2s_1=m\pi/2s_2$ .

\begin{figure}[t]
\epsfxsize9truecm \epsffile{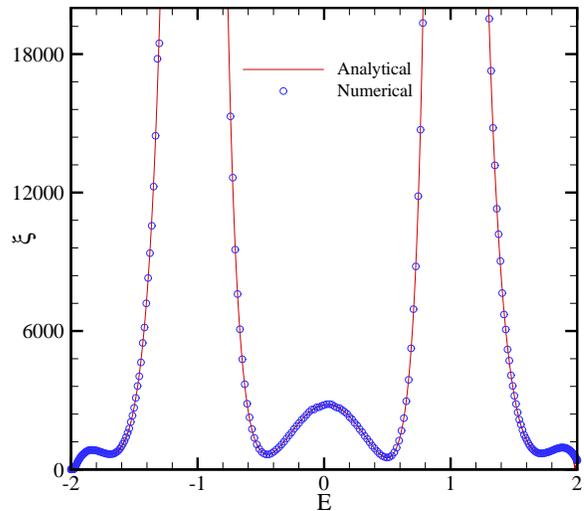} \caption{(Color online) Localization length as a function of energy for binary distribution with $s_1=2, s_2=5$, $ p=q=1/2$ and $U=0.1$. The solid line is the result of weak disorder expansion and circles are data obtained from numerical calculations. States with energies $E=\pm1$ ($k=n\pi/(s_2-s_1)=n\pi/3$) are fully transparent.}\label{fig-xi-2-5}
\end{figure}
\begin{figure}[t]
\epsfxsize9truecm \epsffile{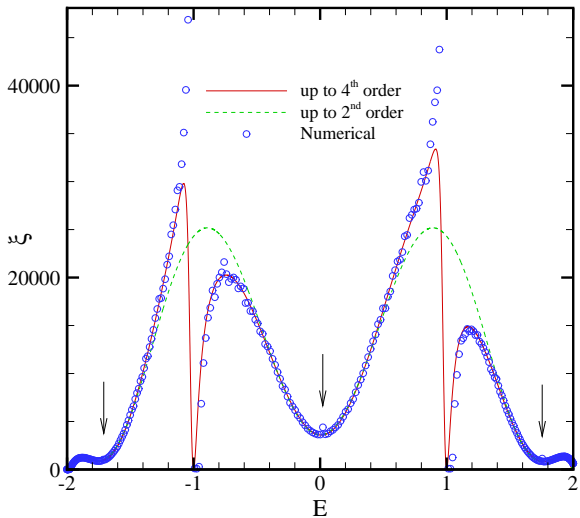} \caption{(Color online) Localization length as a function of energy for binary distribution with $s_1=3, s_2=6$, $ p=q=1/2$ and $U=0.1$. The solid red line shows the analytical result up to $4^{\text{th}}$ order, dashed green line shows the result of weak disorder expansion up to $2^{\text{nd}}$ order and circles are data obtained from numerical calculations. As the numerical results indicate, states with energies $E=\pm1$ are fully transparent. However, since the perturbative expansion fails at these energies so the solid line deviates from numerical results in the neighborhood of $E=\pm1$. The arrows show the position of anomalous energies.}\label{fig-xi-3-6}
\end{figure}
\begin{figure}[t]
\epsfxsize9truecm \epsffile{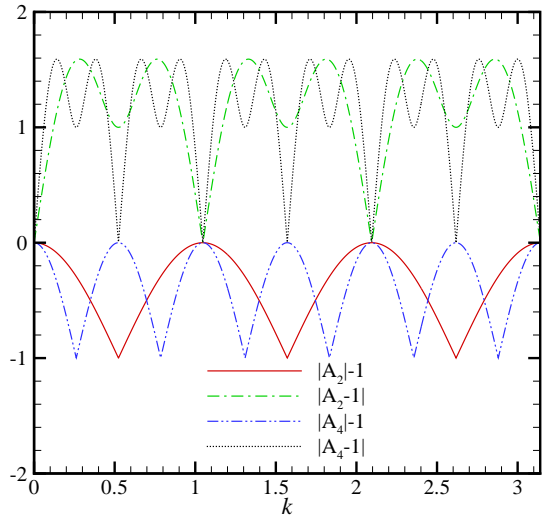} \caption{(Color online) Here we have plotted $|\mathcal{A}_2|-1$ (red), $|\mathcal{A}_2-1|$ (green), $|\mathcal{A}_4|-1$ (blue) and $|\mathcal{A}_4-1|$ (black) as a function of wave vector, for binary distribution with $s_1=3$ and $s_2=6$. The roots of $|\mathcal{A}_2|-1$ and $|\mathcal{A}_2-1|$ correspond to transparent states and band edges respectively and the roots of $|\mathcal{A}_4-1|$ correspond to the anomalous energies (see Fig. \ref{fig-xi-3-6}).} \label{a2a4_3-6}
\end{figure}

\section{Comparison with numerical results}
In order to illustrate these predictions and validate our analytical results, we do the numerical calculation of the Lyapunov exponent, using the standard numerical transfer matrix method, for several distributions of spacing between impurities.

\subsection{Binary distribution}
As a first example, we consider the binary distribution that was mentioned above
\begin{equation}\label{}
p(s)=p \delta_{s,s_1}+ q \delta_{s,s_2},
\end{equation}

The averages that appear in the analytical formula Eq. (\ref{gamma}) are given by
\begin{eqnarray}
&&\langle s\rangle=ps_1+qs_2,\\\label{}
&&\mathcal{A}_2 =pA^{-2s_1}+qA^{-2s_2},\\\label{}
&&\mathcal{A}_4 =pA^{-4s_1}+qA^{-4s_2}.
\end{eqnarray}

Figure \ref{fig-xi-2-5} shows the localization length as a function of energy for a binary distribution where the sequence length takes two different values $s_1=2,s_2=5$ with $p=q=1/2$. The localization length diverges at $E=\pm 1$ which correspond to wave vectors $k=\pi/3$ and $k=2\pi/3$ respectively. As it was discussed above these are transparent states. In this case, the anomalous energy is at the band center.

Figure \ref{fig-xi-3-6} is a similar result for $s_1=3,s_2=6$. Since $s_2-s_1=3$, again we expect $E=\pm 1$ to be transparent states however these energies coincide with energy gaps where the weak disorder expansion fails. Therefore we see deviations from numerical data in the neighborhood of them. In this case, we have three anomalous energies corresponding to $k=\pi/6,\pi/3,2\pi/3$ (see Fig. \ref{a2a4_3-6}). The analytical result up to $2^{\text{nd}}$ order of perturbation \cite{izrailev1999localization} is also included for comparison. As it can be seen, the $2^{\text{nd}}$ order result has considerable deviation from the numerical data in the vicinity of $E=\pm 1$.

\begin{figure}[t]
\epsfxsize9truecm \epsffile{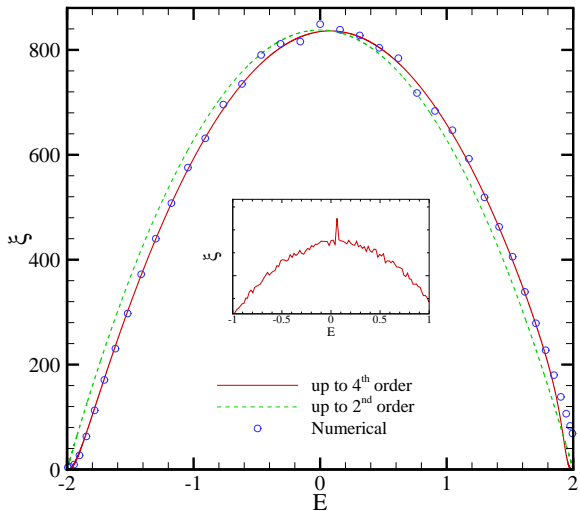} \caption{(Color online) Localization length as a function of energy for exponential distribution with $\mu=0.5$ and $U=0.2$. The solid red line shows the analytical result up to $4^{\text{th}}$ order, the dashed green line shows the result of weak disorder expansion up to $2^{\text{nd}}$ order and circles are data obtained from numerical calculations. The inset shows the numerical result for localization length with $\mu=0.5$ and $U=0.1$ where the band center anomaly can be seen.}\label{xi_exp_0.5}
\end{figure}
\begin{figure}[t]
\epsfxsize9truecm \epsffile{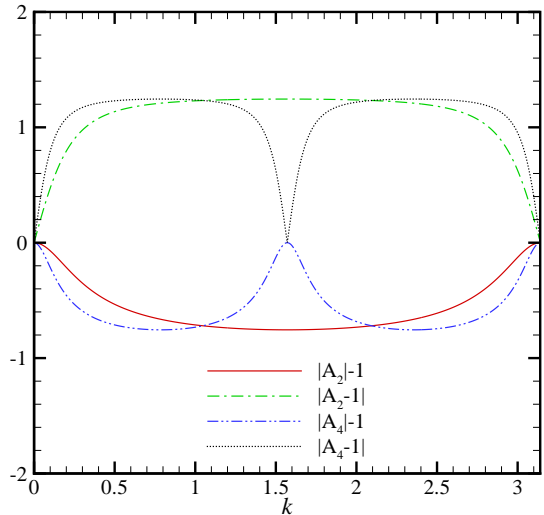} \caption{(Color online) Here we have plotted $|\mathcal{A}_2|-1$ (red), $|\mathcal{A}_2-1|$ (green), $|\mathcal{A}_4|-1$ (blue) and $|\mathcal{A}_4-1|$ (black) as a function of wave vector, for exponential distribution with $\mu=0.5$. The roots of $|\mathcal{A}_2|-1$ and $|\mathcal{A}_2-1|$ correspond to transparent states and band edges respectively and the roots of $|\mathcal{A}_4-1|$ correspond to the anomalous energies (see the inset of Fig. \ref{xi_exp_0.5}).} \label{a2a4_exp}
\end{figure}
\begin{figure}[t]
\epsfxsize9truecm \epsffile{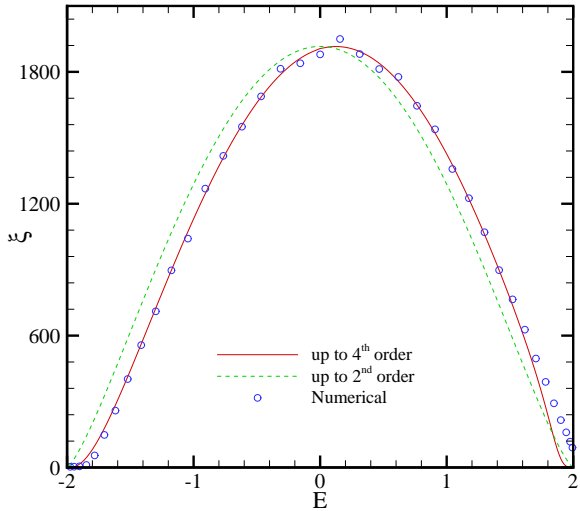} \caption{(Color online) Localization length as a function of energy for power-law distribution with $\alpha=2$ and $U=0.2$. The solid red line shows the analytical result up to $4^{\text{th}}$ order, dashed green line shows the result of weak disorder expansion up to $2^{\text{nd}}$ order and circles are data obtained from numerical calculations.}\label{power-law}
\end{figure}

\subsection{Exponential distribution}

Now let us consider the exponential distribution
\begin{equation}\label{}
 p(s)=\frac{e^{-\mu s}}{e^{\mu}-1},
\end{equation}
where $\mu>0$ and $s\geq 1$. The averages are given by
\begin{eqnarray}
&&\langle s\rangle=\frac{e^{\mu}}{e^{\mu}-1},\\\label{A2exp}
&&\mathcal{A}_2 =\frac{e^{\mu}-1}{A^2 e^{\mu}-1},\\\label{A4exp}
&&\mathcal{A}_4 =\frac{e^{\mu}-1}{A^4 e^{\mu}-1}.
\end{eqnarray}
As it can be seen from Eq. (\ref{A4exp}) and Fig. \ref{a2a4_exp}, in this case, there is only one anomalous energy at the band center, where $A\rightarrow i$, thus $\mathcal{A}_4 \rightarrow 1$. This is shown in Fig. \ref{xi_exp_0.5}. At the band edges there are similar divergencies but with a different degrees because $1-\mathcal{A}_2$ also vanishes.

\subsection{Power-law distribution}

A simple one-dimensional model of a L\'{e}vy glass can be realized by using the power-law distribution for spacing between impurities \cite{zakeri2015localization}
\begin{equation}\label{}
 p(s)=\frac{s^{-(1+\alpha)}}{\zeta(1+\alpha)},
\end{equation}
where $\alpha>0$, $s\geq1$ and $\zeta(z)$ is the Riemann zeta function. For this case we have
\begin{eqnarray}
&&\langle s\rangle=\frac{\zeta(\alpha)}{\zeta(1+\alpha)},\\\label{A2exp}
&&\mathcal{A}_2 =\frac{\text{Li}_{1+\alpha}(A^{-2})}{\zeta(1+\alpha)},\\\label{A4exp}
&&\mathcal{A}_4 =\frac{\text{Li}_{1+\alpha}(A^{-4})}{\zeta(1+\alpha)}.
\end{eqnarray}
where $\text{Li}_{\beta}(z)=\sum_{s=1}^{\infty} z^s s^{-\beta}$. Again an anomalous behavior is expected at the band center because $\text{Li}_{1+\alpha}(1)=\zeta(1+\alpha)$ therefore $\mathcal{A}_4=1$.

Figure \ref{power-law} shows the localization length as a function of energy for the power-law distribution with $\alpha=2$ and $U=0.2$. It is interesting to note that even though the mean spacing between impurities for this power-law distribution ($\langle s\rangle_{\alpha=2}=1.36$) is smaller than the case of exponential distribution in Fig. \ref{xi_exp_0.5} ($\langle s\rangle_{\mu=0.5}=2.54$), the localization length is larger for the power-law distribution.

The power-law distribution with $\alpha \leq 1$ is a peculiar case because the average spacing between consecutive impurities, $\langle s\rangle$, diverges which means zero density of impurities in the thermodynamic limit. In Ref. \cite{zakeri2015localization} it is argued that the Lyapunov exponent should be zero in this range of $\alpha$ but the numerical simulations of Ref. \cite{herrera2019heat} show a nonzero Lyapunov exponent.
Our result shows the explicit dependence on $\langle s\rangle$ in each term of the expansion. The Lyapunov exponent vanishes as the density of the impurities tends to zero. We believe that for a similar reason the Lyapunov exponent will also vanish when $\alpha \leq 1$. However, it should be noted that even though the localization length diverges for this case, the transmission coefficient might vanish in the thermodynamic limit \cite{fernandez2014beyond,falceto2011conductance}. Such states are called anomalously localized.

\section{summary and conclusion}
We have studied electron localization in a one-dimensional lattice consisting of impurities with a given distribution of spacing between them. The model is also applicable to the propagation of classical waves in harmonic chains. Since the potential is correlated one needs to know the correlations in order to obtain the Lyapunov exponent. We introduce a method of obtaining the Lyapunov exponent which does not require the explicit calculation of the correlation functions of the disorder. Our result exhibits the dependence of the Lyapunov exponent on the average spacing between impurities $\langle s\rangle$ explicitly, therefore it is more conclusive in the limit of infinite average spacing compared to previous studies which have used the power spectrum. As $\langle s\rangle$ goes to infinity (the case $\alpha\leq1$ in power-law distribution) the Lyapunov exponent vanishes. Also, our approach allows a systematic calculation of higher orders of perturbative expansion. This allows us to study the anomalous energies where the localization length is enhanced in a narrow window of energy. We show that in addition to the band center anomaly, which occurs in the uncorrelated model, there could be other anomalous energies depending on the distribution function of the spacing between the impurities. We also discuss the range of validity of the perturbation theory and transparent states that might exist in different cases. The method that we introduced in this paper can be applied to other potentials of this type.
\section{acknowledgment}
We would like to acknowledge financial support from the research council of University of Tehran for this research.

\bibliography{myref}

\begin{thebibliography}{22}
\expandafter\ifx\csname natexlab\endcsname\relax\def\natexlab#1{#1}\fi
\expandafter\ifx\csname bibnamefont\endcsname\relax
  \def\bibnamefont#1{#1}\fi
\expandafter\ifx\csname bibfnamefont\endcsname\relax
  \def\bibfnamefont#1{#1}\fi
\expandafter\ifx\csname citenamefont\endcsname\relax
  \def\citenamefont#1{#1}\fi
\expandafter\ifx\csname url\endcsname\relax
  \def\url#1{\texttt{#1}}\fi
\expandafter\ifx\csname urlprefix\endcsname\relax\def\urlprefix{URL }\fi
\providecommand{\bibinfo}[2]{#2}
\providecommand{\eprint}[2][]{\url{#2}}

\bibitem[{\citenamefont{van Rossum and Nieuwenhuizen}(1999)}]{van1999multiple}
\bibinfo{author}{\bibfnamefont{M.~v.} \bibnamefont{van Rossum}}
  \bibnamefont{and} \bibinfo{author}{\bibfnamefont{T.~M.}
  \bibnamefont{Nieuwenhuizen}}, \bibinfo{journal}{Reviews of Modern Physics}
  \textbf{\bibinfo{volume}{71}}, \bibinfo{pages}{313} (\bibinfo{year}{1999}).

\bibitem[{\citenamefont{Ishimaru}(1978)}]{ishimaru1978wave}
\bibinfo{author}{\bibfnamefont{A.}~\bibnamefont{Ishimaru}},
  \emph{\bibinfo{title}{Wave propagation and scattering in random media}},
  vol.~\bibinfo{volume}{2} (\bibinfo{publisher}{Academic press New York},
  \bibinfo{year}{1978}).

\bibitem[{\citenamefont{Maynard}(2001)}]{maynard2001acoustical}
\bibinfo{author}{\bibfnamefont{J.~D.} \bibnamefont{Maynard}},
  \bibinfo{journal}{Reviews of modern physics} \textbf{\bibinfo{volume}{73}},
  \bibinfo{pages}{401} (\bibinfo{year}{2001}).

\bibitem[{\citenamefont{Belitz and Kirkpatrick}(1994)}]{belitz1994anderson}
\bibinfo{author}{\bibfnamefont{D.}~\bibnamefont{Belitz}} \bibnamefont{and}
  \bibinfo{author}{\bibfnamefont{T.}~\bibnamefont{Kirkpatrick}},
  \bibinfo{journal}{Reviews of modern physics} \textbf{\bibinfo{volume}{66}},
  \bibinfo{pages}{261} (\bibinfo{year}{1994}).

\bibitem[{\citenamefont{Barthelemy et~al.}(2008)\citenamefont{Barthelemy,
  Bertolotti, and Wiersma}}]{barthelemy2008levy}
\bibinfo{author}{\bibfnamefont{P.}~\bibnamefont{Barthelemy}},
  \bibinfo{author}{\bibfnamefont{J.}~\bibnamefont{Bertolotti}},
  \bibnamefont{and} \bibinfo{author}{\bibfnamefont{D.~S.}
  \bibnamefont{Wiersma}}, \bibinfo{journal}{Nature}
  \textbf{\bibinfo{volume}{453}}, \bibinfo{pages}{495} (\bibinfo{year}{2008}).

\bibitem[{\citenamefont{Zakeri et~al.}(2015)\citenamefont{Zakeri, Lepri, and
  Wiersma}}]{zakeri2015localization}
\bibinfo{author}{\bibfnamefont{S.~S.} \bibnamefont{Zakeri}},
  \bibinfo{author}{\bibfnamefont{S.}~\bibnamefont{Lepri}}, \bibnamefont{and}
  \bibinfo{author}{\bibfnamefont{D.~S.} \bibnamefont{Wiersma}},
  \bibinfo{journal}{Physical Review E} \textbf{\bibinfo{volume}{91}},
  \bibinfo{pages}{032112} (\bibinfo{year}{2015}).

\bibitem[{\citenamefont{Asatryan and Novikov}(2018)}]{asatryan2018anderson}
\bibinfo{author}{\bibfnamefont{A.~A.} \bibnamefont{Asatryan}} \bibnamefont{and}
  \bibinfo{author}{\bibfnamefont{A.}~\bibnamefont{Novikov}},
  \bibinfo{journal}{Physical Review B} \textbf{\bibinfo{volume}{98}},
  \bibinfo{pages}{235144} (\bibinfo{year}{2018}).

\bibitem[{\citenamefont{Herrera-Gonz{\'a}lez and
  M{\'e}ndez-Berm{\'u}dez}(2019)}]{herrera2019heat}
\bibinfo{author}{\bibfnamefont{I.}~\bibnamefont{Herrera-Gonz{\'a}lez}}
  \bibnamefont{and}
  \bibinfo{author}{\bibfnamefont{J.}~\bibnamefont{M{\'e}ndez-Berm{\'u}dez}},
  \bibinfo{journal}{Physical Review E} \textbf{\bibinfo{volume}{100}},
  \bibinfo{pages}{052109} (\bibinfo{year}{2019}).

\bibitem[{\citenamefont{Izrailev and Krokhin}(1999)}]{izrailev1999localization}
\bibinfo{author}{\bibfnamefont{F.}~\bibnamefont{Izrailev}} \bibnamefont{and}
  \bibinfo{author}{\bibfnamefont{A.}~\bibnamefont{Krokhin}},
  \bibinfo{journal}{Physical review letters} \textbf{\bibinfo{volume}{82}},
  \bibinfo{pages}{4062} (\bibinfo{year}{1999}).

\bibitem[{\citenamefont{Falceto and Gopar}(2010)}]{falceto2011conductance}
\bibinfo{author}{\bibfnamefont{F.}~\bibnamefont{Falceto}} \bibnamefont{and}
  \bibinfo{author}{\bibfnamefont{V.~A.} \bibnamefont{Gopar}},
  \bibinfo{journal}{EPL (Europhysics Letters)} \textbf{\bibinfo{volume}{92}},
  \bibinfo{pages}{57014} (\bibinfo{year}{2010}).

\bibitem[{\citenamefont{Wells~Jr et~al.}(2008)\citenamefont{Wells~Jr, e~Castro,
  and de~Queiroz}}]{wells2008quantum}
\bibinfo{author}{\bibfnamefont{P.}~\bibnamefont{Wells~Jr}},
  \bibinfo{author}{\bibfnamefont{J.~d.} \bibnamefont{e~Castro}},
  \bibnamefont{and}
  \bibinfo{author}{\bibfnamefont{S.}~\bibnamefont{de~Queiroz}},
  \bibinfo{journal}{Physical Review B} \textbf{\bibinfo{volume}{78}},
  \bibinfo{pages}{035102} (\bibinfo{year}{2008}).

\bibitem[{\citenamefont{Iomin}(2009)}]{iomin2009lyapunov}
\bibinfo{author}{\bibfnamefont{A.}~\bibnamefont{Iomin}},
  \bibinfo{journal}{Physical Review E} \textbf{\bibinfo{volume}{79}},
  \bibinfo{pages}{062102} (\bibinfo{year}{2009}).

\bibitem[{\citenamefont{Kappus and Wegner}(1981)}]{kappus1981anomaly}
\bibinfo{author}{\bibfnamefont{M.}~\bibnamefont{Kappus}} \bibnamefont{and}
  \bibinfo{author}{\bibfnamefont{F.}~\bibnamefont{Wegner}},
  \bibinfo{journal}{Zeitschrift f{\"u}r Physik B Condensed Matter}
  \textbf{\bibinfo{volume}{45}}, \bibinfo{pages}{15} (\bibinfo{year}{1981}).

\bibitem[{\citenamefont{Alloatti}(2009)}]{alloatti2008anomalies}
\bibinfo{author}{\bibfnamefont{L.}~\bibnamefont{Alloatti}},
  \bibinfo{journal}{Journal of Physics: Condensed Matter}
  \textbf{\bibinfo{volume}{21}}, \bibinfo{pages}{045503}
  (\bibinfo{year}{2009}).

\bibitem[{\citenamefont{Thouless}(1974)}]{thouless1974electrons}
\bibinfo{author}{\bibfnamefont{D.~J.} \bibnamefont{Thouless}},
  \bibinfo{journal}{Physics Reports} \textbf{\bibinfo{volume}{13}},
  \bibinfo{pages}{93} (\bibinfo{year}{1974}).

\bibitem[{\citenamefont{Balian et~al.}(1983)\citenamefont{Balian, Maynard, and
  rard Toulouse}}]{balian1983ill}
\bibinfo{author}{\bibfnamefont{R.}~\bibnamefont{Balian}},
  \bibinfo{author}{\bibfnamefont{R.}~\bibnamefont{Maynard}}, \bibnamefont{and}
  \bibinfo{author}{\bibfnamefont{G.}~\bibnamefont{rard Toulouse}},
  \emph{\bibinfo{title}{Ill-condensed matter}}, vol.~\bibinfo{volume}{31}
  (\bibinfo{publisher}{World Scientific}, \bibinfo{year}{1983}).

\bibitem[{\citenamefont{Derrida and Gardner}(1984)}]{derrida1984lyapounov}
\bibinfo{author}{\bibfnamefont{B.}~\bibnamefont{Derrida}} \bibnamefont{and}
  \bibinfo{author}{\bibfnamefont{E.}~\bibnamefont{Gardner}},
  \bibinfo{journal}{Journal de Physique} \textbf{\bibinfo{volume}{45}},
  \bibinfo{pages}{1283} (\bibinfo{year}{1984}).

\bibitem[{\citenamefont{Sepehrinia}(2010)}]{sepehrinia2010irrational}
\bibinfo{author}{\bibfnamefont{R.}~\bibnamefont{Sepehrinia}},
  \bibinfo{journal}{Physical Review B} \textbf{\bibinfo{volume}{82}},
  \bibinfo{pages}{045118} (\bibinfo{year}{2010}).

\bibitem[{\citenamefont{Sepehrinia}(2021)}]{sepehrinia2021localization}
\bibinfo{author}{\bibfnamefont{R.}~\bibnamefont{Sepehrinia}},
  \bibinfo{journal}{Physical Review B} \textbf{\bibinfo{volume}{103}},
  \bibinfo{pages}{L020201} (\bibinfo{year}{2021}).

\bibitem[{\citenamefont{Gardner et~al.}(1984)\citenamefont{Gardner, Itzykson,
  and Derrida}}]{gardner1984laplacian}
\bibinfo{author}{\bibfnamefont{E.}~\bibnamefont{Gardner}},
  \bibinfo{author}{\bibfnamefont{C.}~\bibnamefont{Itzykson}}, \bibnamefont{and}
  \bibinfo{author}{\bibfnamefont{B.}~\bibnamefont{Derrida}},
  \bibinfo{journal}{Journal of Physics A: Mathematical and General}
  \textbf{\bibinfo{volume}{17}}, \bibinfo{pages}{1093} (\bibinfo{year}{1984}).

\bibitem[{\citenamefont{Izrailev et~al.}(2001)\citenamefont{Izrailev, Krokhin,
  and Ulloa}}]{izrailev2001mobility}
\bibinfo{author}{\bibfnamefont{F.}~\bibnamefont{Izrailev}},
  \bibinfo{author}{\bibfnamefont{A.}~\bibnamefont{Krokhin}}, \bibnamefont{and}
  \bibinfo{author}{\bibfnamefont{S.}~\bibnamefont{Ulloa}},
  \bibinfo{journal}{Physical Review B} \textbf{\bibinfo{volume}{63}},
  \bibinfo{pages}{041102} (\bibinfo{year}{2001}).

\bibitem[{\citenamefont{Fern{\'a}ndez-Mar{\'\i}n
  et~al.}(2014)\citenamefont{Fern{\'a}ndez-Mar{\'\i}n, M{\'e}ndez-Berm{\'u}dez,
  Carbonell, Cervera, S{\'a}nchez-Dehesa, and Gopar}}]{fernandez2014beyond}
\bibinfo{author}{\bibfnamefont{A.~A.} \bibnamefont{Fern{\'a}ndez-Mar{\'\i}n}},
  \bibinfo{author}{\bibfnamefont{J.}~\bibnamefont{M{\'e}ndez-Berm{\'u}dez}},
  \bibinfo{author}{\bibfnamefont{J.}~\bibnamefont{Carbonell}},
  \bibinfo{author}{\bibfnamefont{F.}~\bibnamefont{Cervera}},
  \bibinfo{author}{\bibfnamefont{J.}~\bibnamefont{S{\'a}nchez-Dehesa}},
  \bibnamefont{and} \bibinfo{author}{\bibfnamefont{V.}~\bibnamefont{Gopar}},
  \bibinfo{journal}{Physical Review Letters} \textbf{\bibinfo{volume}{113}},
  \bibinfo{pages}{233901} (\bibinfo{year}{2014}).

\end{thebibliography}
\bibliographystyle{apsrev}

\end{document}